\documentclass[12pt,a4paper,fleqn]{article}

\usepackage[articletitle=true,super=false,chaptertitle=true,maxauthors=0]{achemso}
\usepackage{amsmath}
\usepackage{amssymb}
\usepackage{mathtools}
\usepackage{graphicx}
\usepackage[american]{babel}
\usepackage{here}
\usepackage{braket}
\usepackage{hyperref}
\usepackage{bbold}
\usepackage{enumitem}
\usepackage{csquotes}

\begin{document}

\begin{center}

{\LARGE\bf
    eCAS: An Orbital Embedding Method for Quantum Computing
}

\vspace{1cm}

{\large
Anna Brünisholz$^{a,}$\footnote{ORCID: 0009-0005-9724-9439} and
Markus Reiher$^{a,}$\footnote{Corresponding author; e-mail: mreiher@ethz.ch; ORCID: 0000-0002-9508-1565}
}\\[4ex]

$^{a}$ Department of Chemistry and Applied Biosciences, ETH Zurich, Vladimir-Prelog-Weg 2,\\
8093 Zurich, Switzerland

October 30, 2026

\vspace{.43cm}

\textbf{Abstract}
\end{center}
\vspace*{-.41cm}
{\small
Future fault-tolerant quantum computers will allow us to calculate electronic structures of atomistic systems. However, these systems will be intrinsically limited in size, requiring embedding theories for many systems because of the unfavorable scaling of the number of terms describing the electron--electron interaction in the second-quantized electronic Hamiltonian.
Quantum embedding theories address this challenge by partitioning a quantum system into a region of interest, the impurity, and its (larger) complement, the environment.
However, currently available quantum embedding methods place restrictive demands on near-future quantum computers, such as the need to measure reduced density matrices or the need for repeated evaluations. 
Here, we introduce the new orbital embedding method eCAS that steers clear of these pitfalls by constructing a multi-CAS problem, the active spaces of which are localized and can be solved independently of each other with any quantum algorithm.
This also allows for the development of comparatively simple gradient expressions.
eCAS leverages the Schmidt decomposition to construct initial impurity and bath orbitals and augments them with localized environment orbitals, based on a mutual correlation criterion.
The mutual correlation criterion makes eCAS systematically improvable without the need for any iteratively determined quantities.
}

\section{Introduction}
\label{sec:introduction}
With the advent of quantum computing, quantum embedding theories~\cite{neug14,chan16,mill19,gagl26,yang23,ratn20} have become central to quantum computing applications~\cite{aspu11,aspu19,chan20,troy22}.
Without embedding, truncation, or downfolding~\cite{lowd62,kowa19,jong20} none but the smallest atomic systems could be treated on a quantum computer.

Quantum embedding theories enable a system to be partitioned into subsystems that are described at different levels of approximation.
The impurity subsystem is typically the one that shall receive the most accurate quantum description and shall therefore be subjected to quantum computation, while the environment subsystem can be described by traditional quantum chemical approaches handled on a classical computer.
Many chemical systems, such as enzymes, naturally partition into an impurity subsystem (the active site) and the more subordinate environment subsystem (distant side chains).

Various quantum embedding methods have already been considered for quantum computations~\cite{gall22_b,dunc26,cove24,tave23,gagl25_b,loug20,dunc20,rung21,arit23,elsä25,gall22_c}.
Most prominent among them is the density matrix embedding theory (DMET)~\cite{rubi16,boot21,yama21,ramo22,lvdi22,mont22,daha23}, which leverages the Schmidt decomposition of a system-wide reference wave function to construct bath orbitals for the chosen impurities~\cite{chan12,chan13,chan17}.
However, because this construction is performed independently for each impurity, no global wave function exists.
While observables local to the impurities can still be evaluated from the impurity Hamiltonians, the absence of a global wave function prevents direct access to global observables, most notably the system energy.
Instead, global observables have to be calculated from partitioned one- and two-body reduced density matrices (1- and 2-RDMs) in a democratic partitioning of the expectation values~\cite{chan17}.
This necessitates the calculation of RDMs on the quantum computer.
Reduced density matrices can be calculated by quantum algorithms, 
albeit at tremendous costs, which would render a quantum advantage unlikely because efficient algorithms primarily provide energies.
DMET is therefore restricted to quantum algorithms that establish an explicit approximation to the many-electron wave function such as the variational quantum eigensolver (VQE)~\cite{obri14,parr21} or sample-based quantum diagonalization~\cite{mezz25}, which are designed to utilize noisy small-scale quantum computers.
These algorithms stand in contrast to quantum algorithms that directly measure energies, such as quantum phase estimation~\cite{kita95,lloy99,aspu09}.
Furthermore, the absence of a global wave function makes the derivation of analytical gradients harder and computationally more expensive~\cite{chan23,merz26}.
Additionally, DMET requires two nested self-consistency loops: one to converge a global chemical potential and one to converge impurity-specific one-body potentials.
The impurity problem must therefore be solved repeatedly with the impurity method.
For quantum computing applications this is problematic, as initialization of quantum computations and switching between quantum and classical hardware is slow~\cite{smit20}.

Advanced embedding methods have been devised that build on DMET, including density embedding theory (DET)~\cite{duke14,scus14} and bootstrap embedding~\cite{voor16,voor23_b,voor25}. They improve convergence of the self-consistency cycles, but still require them.
An exception is an embedding approach developed by Nusspickel and Booth~\cite{boot22,lvdi25,mott26}.
This method does not have an official name; in its `Vayesta'~\cite{boot22,boot23} implementation it is referred to as EWF, and we adopt this name here.
EWF does not require iterative convergence and works in a one-shot procedure.
This is because EWF does not use local one-body potentials to improve the bath-construction and instead adds environment orbitals to the bath.
The environment orbitals are constructed as impurity-specific natural-orbital-like orbitals called bath natural orbitals.
Their occupation numbers provide an ordering of the environment orbitals, which makes EWF systematically improvable,
a highly desirable feature.
However, EWF also does not possess a global wave function and calculates the system-wide energy from projected amplitudes.

Other quantum embedding theories that have been used for quantum computing, such as dynamical mean field theory~\cite{voll89,mull89,kotl92,jarr92,kotl97,loug20,dunc20,rung21,arit23,elsä25} or the localized active space self-consistent field (LASSCF) method~\cite{gagl19,gagl20,gagl22_b,gagl25_b}, often require iterative evaluations of the impurity problem.
Quantum embedding theories that are non-iterative, such as Huzinaga embedding~\cite{kall16,feld23,tave23,cove24,reih25}
or quantum defect embedding theory~\cite{gall22,gall22_c}, perform system-wide calculations with density functional theory (DFT) or a Green's function method.
Unfortunately, such DFT-rooted approaches require approximate density functionals which can deteriorate the accuracy of the quantum description, annihilating the key advantage of quantum computations to deliver error-controlled many-body energies. 
Moreover, Green's function methods are computationally more expensive and therefore not suitable for the description of large molecular systems.

Hence, none of the available quantum embedding theories fulfills all our points from our wishlist for a quantum embedding theory, namely to be agnostic with respect to the quantum algorithm, to be non-iterative, to deliver a global wave function and to be systematically improvable.
For this reason, we introduce our new orbital embedding method eCAS (short for embedded complete active space method), which prepares and selects orbitals for an approach based on many complete active spaces (CASs).
eCAS constructs initial local active spaces, which it then improves upon with a quantum-information-based measure.
Similarly to methods like DMET and EWF (and related to the atomic valence active space method (AVAS)~\cite{kniz17}), eCAS constructs initial impurity and bath orbitals using a Schmidt decomposition.
It does so jointly for all impurities, such that we obtain one Schmidt basis only.
As in EWF, orbitals are then transferred from the environment to the bath.
We decide which orbitals to transfer based on a mutual correlation criterion~\cite{evan25}, which makes the method systematically improvable and yields a quality-measure estimate.
The different impurities and baths form a multi-CAS problem where each active space can be solved independently of the others.
eCAS can be used for one-shot procedures that do not require the calculation of 1- and 2-RDMs with the impurity method. Moreover, an easy access to an analytical gradient is guaranteed.

This work is organized as follows: We begin with an outline of the underlying theory of eCAS, including a discussion of the Schmidt decomposition and the mutual correlation criterion.
We then introduce and elaborate on the eCAS algorithm.
Results of eCAS are compared to results of EWF and DMET.
We also show the efficacy of the quality measure.
Finally, we summarize our conclusions and discuss future developments of eCAS.

\section{Theory}
We denote the full system with the subscript `AB' and partition it into an impurity (subscript `A') and an environment (subscript `B').
The impurity is then described with an impurity electronic-structure model, while the environment is described with an environment electronic-structure model.
eCAS constructs bath orbitals that capture all mean-field entanglement between impurity and environment using a Schmidt decomposition~\cite{schm07,scho11} of the mean-field reference wave function.
Impurity and bath orbitals form the extended impurity $\mathcal{A}$.
The remaining orbitals form the reduced environment $\mathcal{B}$, which does not contain the bath orbitals, despite them also being part of the initial environment subspace.
In the next step, the extended impurity is augmented by orbitals transferred from the reduced environment $\mathcal{B}$ to it.
This is accomplished based on a mutual correlation criterion.
Finally, the augmented extended impurity becomes the active orbital space in a CAS calculation.
Despite eCAS relying on an initial mean-field wave function, this does not mean that the mean-field calculation has to be a good reference calculation.
As long as the subspace spanned by the impurity, bath and transferred environment orbitals is able to capture the multi-reference behavior of the exact ground state wave function, and the impurity method is a multi-reference method, strongly correlated systems can still be adequately described.

\subsection{Selecting Basis and Extended Impurity}
To construct the extended impurity, a Schmidt decomposition of the restricted Hartree--Fock (HF) wave function is performed.
The HF wave function is constructed from an orthonormal set of $d_{AB}$ spatial orbitals $\{\phi_{AB, i}^\mathrm{(HF)}\}_{i=1}^{d_{AB}}$, $n_{AB}$ of which are doubly occupied.
We select the initial impurity subspace for the Schmidt decomposition from $d_A$ orthonormalized orbitals of the impurity atoms or orbitals, $\{\phi_{A, i}\}_{i=1}^{d_A}$.
The Schmidt decomposition is then performed using the projected overlap matrices $\mathbf{S}$ and $\mathbf{Q}$
\begin{subequations}
\begin{align}
     S_{pq} &= \braket{\phi_{AB, p}^\mathrm{(HF)} | \Hat{P}_A | \phi_{AB, q}^\mathrm{(HF)}}, \quad p, q \in \{1, \dots, n_{AB}\} \\
     Q_{pq} &= \braket{\phi_{AB, p}^\mathrm{(HF)} | \Hat{P}_A | \phi_{AB, q}^\mathrm{(HF)}}, \quad p, q \in \{n_{AB}+1, \dots, d_{AB}\} \\
     \Hat{P}_A &= \sum_{i=1}^{d_{A}} \ket{\phi_{A, i}} \bra{\phi_{A, i}},
\end{align}
\end{subequations}
which are then diagonalized
\begin{subequations}
\begin{align}
    \mathbf{S} &= \mathbf{V} \mathrm{diag}(\lambda_i) \mathbf{V}^\dagger, \quad i\in\{1, \dots, n_{AB}\} \\
    \mathbf{Q} &= \mathbf{U} \mathrm{diag}(\sigma_i) \mathbf{U}^\dagger, \quad i\in\{n_{AB}+1, \dots, d_{AB}\}.
\end{align}
\end{subequations}
It holds that $\lambda_i \in [0, 1]$, $\sigma_i \in [0, 1]$ and at least $n_{AB}-d_A$ eigenvalues $\lambda_i$ and $d_{AB}-n_{AB}-d_A$ eigenvalues $\sigma_i$ are equal to zero.
This is because $\Hat{P}_A$ is a projection operator with rank $d_A$.
For simplicity, we from now on assume that none of the eigenvalues $\{\lambda_i\}_{i=1}^{n_{AB}}$ or $\{\sigma_i\}_{i=n_{AB}+1}^{d_{AB}}$ are equal to one.
As a corollary, exactly $n_{AB}-d_A$ eigenvalues $\lambda_i$ and exactly $d_{AB}-n_{AB}-d_A$ eigenvalues $\sigma_i$ are zero.
Applying $\mathbf{V}^\dagger$ and $\mathbf{U}^\dagger$ to the molecular orbitals (MOs) yields the rotated MOs $\{\Tilde{\phi}_{AB, i}\}_{i=1}^{d_{AB}}$.
They can either be projected into the impurity
\begin{align}
    \ket{\Tilde{\phi}_{A, i}} =\begin{cases}
          \frac{\sum_{p=1}^{n_{AB}} \Hat{P}_A V_{pi}^* \ket{\phi_{AB, p}^\mathrm{(HF)}} }{\sqrt{\lambda_i}}, & \text{if $i \in \{1, 2, \dots, n_{AB}\}$ }\\
    \frac{\sum_{p=n_{AB}+1}^{d_{AB}} \Hat{P}_A U_{pi}^* \ket{\phi_{AB, p}^\mathrm{(HF)}} }{\sqrt{\sigma_i}}, & \text{if $i \in \{n_{AB}+1, n_{AB}+2, \dots, d_{AB}\}$}
    \end{cases}
\end{align}
or into the environment
\begin{align}
    \ket{\Tilde{\phi}_{B, i}} = \begin{cases}
    \frac{\sum_{p=1}^{n_{AB}} \Hat{P}_B V_{pi}^* \ket{\phi_{AB, p}^\mathrm{(HF)}} }{\sqrt{1-\lambda_i}}, & \text{if $i \in \{1, 2, \dots, n_{AB}\}$} \\
    \frac{\sum_{p=n_{AB}+1}^{d_{AB}} \Hat{P}_B U_{pi}^* \ket{\phi_{AB, p}^\mathrm{(HF)}} }{\sqrt{1-\sigma_i}}, & \text{ if $i \in \{n_{AB}+1, n_{AB}+2, \dots, d_{AB}\}$}.
    \end{cases}
\end{align}
The denominators in the construction of the rotated MO basis $\{\Tilde{\phi}_{AB, i}\}_{i=1}^{d_{AB}}$ are to be understood as normalization factors.
The projection operator into the environment is
\begin{align}
    \Hat{P}_B = \Hat{1}-\Hat{P}_A.
\end{align}
Non-zero rotated impurity MOs are the same for occupied and virtual orbitals, as they are projected into the same space which fully lies in the MO span.
We can thus remove the $d_A$ rotated impurity MOs constructed from virtual orbitals.
Similarly, non-zero rotated environment MOs constructed from virtual orbitals are redundant.
These removals yield a complete orthonormal basis of rotated orbitals $\{\Tilde{\phi}_{AB, i}\}_{i=1}^{d_{AB}}$.
Additionally, we exploit that rotated MOs with $\lambda_i = 0$, or with $\sigma_i = 0$, are part of degenerate subspaces.
This is used to localize the environment orbitals.

The Schmidt decomposition of an \emph{unrestricted} HF wave function, where $n_{AB}$ spin orbitals are occupied, is then
\begin{subequations}
\begin{align}
    \ket{\Psi_{AB}^\mathrm{(HF)}} &= \prod_{p=1}^{n_{AB}} \Hat{a}^\dagger(\phi_{AB, p}^\mathrm{(HF)}) \ket{\mathrm{vac}} \\
    &= \prod_{i=1}^{n_{AB}} \Hat{a}^\dagger \left (\sum_{p=1}^{n_{AB}} V_{pi}^* (\Hat{P}_A + \Hat{P}_B) \phi_{AB, p}^\mathrm{(HF)} \right) \ket{\mathrm{vac}} \\
    &= \prod_{i=1}^{n_{AB}} \left( \sqrt{\lambda_i} \Hat{a}^\dagger(\Tilde{\phi}_{A, i}) + \sqrt{1-\lambda_i} \Hat{a}^\dagger(\Tilde{\phi}_{B, i}) \right) \ket{\mathrm{vac}}. \label{eq:Schmidttriv}
\end{align}
\end{subequations}
Here, we exploit the invariance of observables upon unitary rotation of the orbitals.
$\Hat{a}^\dagger(\phi)$ creates an electron in spin orbital $\phi$.
Because $n_{AB}-d_A$ eigenvalues $\lambda_i$ are zero, $\{\Tilde{\phi}_{B, i}\}_{i=1}^{n_{AB}}$ can be further categorized into core orbitals $\{\Tilde{\phi}_{\mathrm{core}, i}\}_{i=1}^{n_{AB}-d_A}$ and bath orbitals $\{\Tilde{\phi}_{\mathrm{bath}, i}\}_{i=1}^{d_A}$
\begin{align}
    \ket{\Psi_{AB}^\mathrm{(HF)}} =& \prod_{i=1}^{n_{AB}-d_{A}} \Hat{a}^\dagger(\Tilde{\phi}_{\mathrm{core}, i})
     \prod_{i=1}^{d_A} \left( \sqrt{\lambda_i} \Hat{a}^\dagger(\Tilde{\phi}_{A, i}) + \sqrt{1-\lambda_i} \Hat{a}^\dagger(\Tilde{\phi}_{\mathrm{bath}, i}) \right) \ket{\mathrm{vac}}. \label{eq:decomp}
\end{align}
Hence, impurity and bath orbitals are pairwise entangled.
We can now construct the complete and orthonormal Schmidt basis $\{\phi_{\mathcal{AB}, i}\}_{i=1}^{d_{AB}}$, as obtained from a \emph{restricted} HF calculation:
\begin{itemize}
    \item $\{\phi_{\mathcal{A}, i}\}_{i=1}^{2d_A}$:
    The extended impurity orbitals.
    They are formed from the union of $\{\Tilde{\phi}_{A, i}\}_{i=1}^{n_{AB}}$ with $0 < \lambda_i \leq 1$ (the $d_A$ impurity orbitals $\{\Tilde{\phi}_{A, i}\}_{i=1}^{d_{A}}$) and $\{\Tilde{\phi}_{B, i}\}_{i=1}^{n_{AB}}$ with $0 < \lambda_i \leq 1$ (the $d_A$ bath orbitals $\{\Tilde{\phi}_{\mathrm{bath}, i}\}_{i=1}^{d_{A}}$).
    These spatial orbitals contain $2d_{A}$ electrons.
    \item $\{\phi_{\mathcal{B}, i}\}_{i=1}^{d_{AB} - 2d_A}$:
    The reduced environment orbitals.
    They can be further categorized into occupied environment orbitals $\{\Tilde{\phi}_{\mathrm{core}, i}\}_{i=1}^{n_{AB}-d_A}$ and virtual environment orbitals $\{\Tilde{\phi}_{\mathrm{vir}, i}\}_{i=1}^{d_{AB}-n_{AB}-d_A}$.
    The occupied environment orbitals are the $n_{AB} -d_A$ rotated occupied environment MOs $\{\Tilde{\phi}_{B, i}\}_{i=1}^{n_{AB}}$ where $\lambda_i = 0$.
    Conversely, the virtual environment orbitals are the $d_{AB} - n_{AB} -d_A$ rotated virtual environment MOs $\{\Tilde{\phi}_{B, i}\}_{i=n_{AB}+1}^{d_{AB}}$ where $\sigma_i = 0$.
\end{itemize}

In order to treat more than one impurity, we subdivide the extended impurity orbitals $\{\phi_{\mathcal{A}, i}\}_{i=1}^{2d_A}$ into $N_\mathrm{imp}$ smaller, non-overlapping extended impurities $\{\phi_{\mathcal{A}_x, i}\}_{i=1}^{2d_{A_x}}$.
Thereby, $d_{A_x}$ is chosen from the chemical structure of the molecule.
The rotation matrices $\mathbf{V}$ and $\mathbf{U}$ are still constructed as if the impurity were not subdivided.
The assignment of impurity--bath pairs $\Tilde{\phi}_{A, i}, \Tilde{\phi}_{\mathrm{bath}, i}$ to different impurities is accomplished based on the minimization of the mutual correlation (see next section).

\subsection{Subdividing and Augmenting the Extended Impurity} \label{sec:augact}
The partitioning of the extended impurity $\mathcal{A}$ into separate smaller extended impurities $\{\mathcal{A}_x\}_{x=1}^{N_\mathrm{imp}}$ is carried out such that the mutual correlation $\mathcal{M}$~\cite{evan25} 
between different extended impurities is minimal.
Here, the mutual correlation serves as a computationally less expensive stand-in for the mutual information~\cite{soly03,whit06,reih12,reih19}.
In its spin-free form, the mutual correlation between orbital $i$ and orbital $j$ is obtained as
\begin{align}
    \mathcal{M}_{ij} &= (\Lambda^2)_{ijjj} + \frac{1}{2} (\Lambda^2)_{iijj} + (\Lambda^2)_{ijij} + (\Lambda^2)_{iiji},
\end{align}
where the individual terms are prepared from the two-body cumulant~\cite{pitt26}
\begin{align}
    \begin{split}
    (\Lambda^2)_{ijkl} &= |\lambda_{I\uparrow J\uparrow K\uparrow L\uparrow}|^2 +  |\lambda_{I\uparrow J\downarrow K\uparrow L\downarrow}|^2 + |\lambda_{J\uparrow I\downarrow K\uparrow L\downarrow}|^2 \\
    &\qquad  + |\lambda_{I\uparrow J\downarrow L\uparrow K\downarrow}|^2 + |\lambda_{J\uparrow I\downarrow L\uparrow K\downarrow}|^2 + |\lambda_{I\downarrow J\downarrow K\downarrow L\downarrow}|^2
    \end{split}
\end{align}
and capital letters indicate spin orbitals.
The two-body cumulant itself is defined as those contributions to the 2-RDM that cannot be expressed with the 1-RDMs, such that
\begin{align}
    \lambda_{IJKL} = \gamma_{IJKL} - (\gamma_{IK}\gamma_{JL} - \gamma_{IL}\gamma_{JK}).
\end{align}
1- and 2-RDM are defined as
\begin{subequations}
\begin{align}
    \gamma_{ij} &= \braket{\Psi|\Hat{a}^\dagger(\phi_i)\Hat{a}(\phi_j)|\Psi} \\
    \gamma_{ijkl} &= \braket{\Psi|\Hat{a}^\dagger(\phi_i)\Hat{a}^\dagger(\phi_j)\Hat{a}(\phi_l)\Hat{a}(\phi_k)|\Psi}.
\end{align}
\end{subequations}
For singlet states, spin-dependent density matrices can be obtained from the spin-summed quantities~\cite{mukh10}.
If we assume that the size of the $\mathcal{A}_x$ is pre-determined, we can partition $\mathcal{A}$ by minimizing the sum of the inter-impurity mutual correlation
\begin{align} \label{eq:partA}
    \min \left( \sum_{x=1}^{N_\mathrm{imp}}\sum_{y > x} \sum_{i\in \mathcal{A}_x} \sum_{j\in \mathcal{A}_y} \mathcal{M}_{ij}\right).
\end{align}
Here, we optimize over the partitioning of $\{\mathcal{\phi}_{\mathcal{A}, i}\}_{i=1}^{2d_A}$ into the smaller sub-impurities $\mathcal{A}_1, \dots, \mathcal{A}_{N_{\mathrm{imp}}}$ under two constraints. 
a) the size of the impurities is fixed ($|\mathcal{A}_x| = 2d_{A_x}$ for all $x \in \{1, \dots, N_{\mathrm{imp}}\}$) and b) impurity--bath pairs are always kept together.

The mutual correlation is also used within eCAS to augment the extended impurities.
This is necessary because the bath orbitals were constructed without considering electron correlation.
We transfer the orbital $\phi_{\mathcal{B}, j}$ from the reduced environment $\mathcal{B}$ to the extended impurity $\mathcal{A}_x$ if
\begin{align}
    \max_{i\in\mathcal{A}_x} \mathcal{M}_{i, j\in\mathcal{B}} \geq \kappa.
\end{align}
If the threshold $\kappa$ is reached for more than one extended impurity $\mathcal{A}_x$, $\phi_{\mathcal{B}, j}$ is assigned to the extended impurity with which it has the largest mutual correlation.
Orbitals transferred from $\mathcal{B}$ to $\mathcal{A}$ are not used to recruit other orbitals.

Finally, we note that one may also apply the concept of mutual information~\cite{soly03,whit06,reih12,reih19} instead of mutual correlation.
We defer the elaboration of this option to future work.

\subsection{Estimating the Quality of eCAS}
As an estimate of the quality of the embedding ansatz, we use the inter-extended-impurity and extended-impurity--environment mutual correlation~\cite{evan25} $\mathcal{Q}$
\begin{subequations}
\begin{align}
    \mathcal{Q} =& \sum_{X<Y} \mathcal{M}_{XY} + \sum_{X<Y<Z} \mathcal{M}_{XYZ} + \sum_{X<Y<Z<W} \mathcal{M}_{XYZW} \\
    \begin{split}
    \mathcal{M}_{XY} =&  \sum_{p\in X} \sum_{qrs\in Y} (\Lambda^2)_{pqrs} + \frac{1}{2} \sum_{pq\in X} \sum_{rs \in Y} (\Lambda^2)_{pqrs} \\
    & + \sum_{pr\in X} \sum_{qs \in Y} (\Lambda^2)_{pqrs}
    + \sum_{pqr\in X} \sum_{s\in Y} (\Lambda^2)_{pqrs}
    \end{split} \\
    \begin{split}
    \mathcal{M}_{XYZ} =&  \sum_{p\in X} \sum_{q\in Y} \sum_{rs \in Z} \left( (\Lambda^2)_{pqrs} + 2(\Lambda^2)_{prqs} \right) \\
     & + \sum_{p\in Y} \sum_{q\in Z} \sum_{rs \in X} \left( (\Lambda^2)_{pqrs} + 2(\Lambda^2)_{prqs} \right) \\
     & + \sum_{p\in X} \sum_{q\in Z} \sum_{rs \in Y} \left( (\Lambda^2)_{pqrs} + 2(\Lambda^2)_{prqs} \right)
    \end{split} \\
    \mathcal{M}_{XYZW} =& 6 \sum_{p\in X} \sum_{q\in Y} \sum_{r \in Z} \sum_{s \in W} (\Lambda^2)_{pqrs}.
\end{align}
\end{subequations}
$X, Y, Z$ and $W$ represent the different extended impurities $\{ \mathcal{A}_x \}_{x=1}^{N_\mathrm{imp}}$, as well as the reduced environment $\mathcal{B}$.

In the case that the correlated wave function can be represented as a product wave function with the same orbital groupings as represented with eCAS, it will hold that $\mathcal{Q} = 0$.
$\mathcal{Q}$ does not consider any correlation within the subsystems, only between them.
This allows us to disregard any correlation in the environment, as we assume that it is irrelevant for observables local to the impurities.

\subsection{Constructing the Multi-CAS Hamiltonian}
We now construct the multi-CAS Hamiltonian $\Hat{H}_{{\mathcal{A}}_1, \dots, {\mathcal{A}}_{N_\mathrm{imp}}\mathcal{B}}$.
Therein, the different augmented extended impurities take the role of active spaces, which are only coupled via a mean field
\begin{subequations} \label{eq:DMETmult}
\begin{align}
    \Hat{H}_{{\mathcal{A}}_1, \dots, {\mathcal{A}}_{N_\mathrm{imp}}\mathcal{B}} =&  E_{\mathrm{nn}, AB} + E_{\mathrm{core}} + \sum_{x=1}^{N_\mathrm{imp}} \Hat{H}_{\mathcal{A}_x} - E_{\mathrm{dc}} \\
    E_{\mathrm{nn}, AB} =& \sum_{I<J} \frac{Z_I Z_J}{|\mathbf{R}_I - \mathbf{R}_J|} \label{eq:Enn} \\
    E_{\mathrm{core}} =& \mathrm{tr}({\boldsymbol{\gamma}_{\mathcal{B}} \mathbf{h}_{0, AB}}) + E_\mathrm{ee}[\boldsymbol{\gamma}_{\mathcal{B}}] \\
    \begin{split}
    \Hat{H}_{\mathcal{A}_x} =& \sum_{ij \in \mathcal{A}_x} h_{\mathcal{A}_x, ij} \Hat{a}^\dagger (\phi_{\mathcal{A}_x, i}) \Hat{a}(\phi_{\mathcal{A}_x, j}) \\
    & + \frac{1}{2} \sum_{ijkl \in \mathcal{A}_x} g_{\mathcal{A}_x, ijkl} \Hat{a}^\dagger(\phi_{\mathcal{A}_x, i})\Hat{a}^\dagger(\phi_{\mathcal{A}_x, j}) \Hat{a}(\phi_{\mathcal{A}_x, l}) \Hat{a}(\phi_{\mathcal{A}_x, k})
    \end{split} \\
    E_{\mathrm{dc}} =& \sum_{x=1}^{N_\mathrm{imp}}\sum_{y>x} E_{\mathrm{ee}}[\boldsymbol{\gamma}_{\mathcal{A}_x} + \boldsymbol{\gamma}_{\mathcal{A}_y} ] - E_{\mathrm{ee}}[\boldsymbol{\gamma}_{\mathcal{A}_x}] -E_{\mathrm{ee}}[\boldsymbol{\gamma}_{\mathcal{A}_y}].
\end{align}
\end{subequations}
The nuclear--nuclear repulsion energy $E_{\mathrm{nn}, AB}$ depends on the charge $Z$ and location $\mathbf{R}$ of the nuclei.
The core energy is obtained from the original system-wide restricted HF calculation, where, in the atomic orbital representation
\begin{subequations}
\begin{align}
\begin{split}
    h_{0, AB, ij} &= -\frac{1}{2} \braket{\chi_{AB, i}(\mathbf{r})|\nabla^2|\chi_{AB, j}(\mathbf{r})} \\
    &\quad - \sum_{I=1} \braket{\chi_{AB, i}(\mathbf{r})|\frac{Z_I}{|\mathbf{r}- \mathbf{R}_I|} | \chi_{AB, j}(\mathbf{r})}, \label{eq:coreH}
\end{split} \\
    \mathbf{V}[\boldsymbol{\gamma}_{AB}] &= \mathbf{J}[\boldsymbol{\gamma}_{AB}] + \mathbf{K}[\boldsymbol{\gamma}_{AB}] \\
    \begin{split}
        J_{ij}[\boldsymbol{\gamma}_{AB}] &= \sum_{k, l}^{d_{AB}} \gamma_{AB, kl} \braket{\chi_{AB, i} (\mathbf{r}_1)\chi_{AB, k} (\mathbf{r}_2)| \frac{1}{|\mathbf{r}_1 - \mathbf{r}_2|}|\chi_{AB, l} (\mathbf{r}_2) \chi_{AB, j} (\mathbf{r}_1)}
    \end{split} \\
    \begin{split}
    K_{ij}[\boldsymbol{\gamma}_{AB}] &= -\frac{1}{2} \sum_{k, l}^{d_{AB}} \gamma_{AB, kl} \\
    &\quad\quad \cdot \braket{\chi_{AB, i} (\mathbf{r}_1)\chi_{AB, k} (\mathbf{r}_2) | \frac{1}{|\mathbf{r}_1 - \mathbf{r}_2|} | \chi_{AB, j} (\mathbf{r}_2) \chi_{AB, l} (\mathbf{r}_1)}.
    \end{split}
\end{align}
\end{subequations}
$\{\chi_{AB, i}\}_{i=1}^{d_{AB}}$ is the atomic orbital basis set.
The HF electron--electron repulsion energy is
\begin{align}
    E_{\mathrm{ee}}[\boldsymbol{\gamma}_{AB}] &= \frac{1}{2} \mathrm{tr}((\mathbf{J} + \mathbf{K})\boldsymbol{\gamma}_{AB}).
\end{align}
The active space Hamiltonians are obtained while also considering the mean-field coupling to other impurities
\begin{subequations}
\begin{align}
    h_{\mathcal{A}_x, ij} &= \sum_{pq \in AB} C_{pi}^*  h_{(AB)_x, pq} C_{qj} \\
    g_{\mathcal{A}_x, ijkl} &= \sum_{pqrs \in AB} C_{pi}^* C_{qj}^* g_{AB, pqrs} C_{rk}C_{sl} \\
    h_{(AB)_x, ij} &= h_{0, AB, ij} + V_{ij}\left[\boldsymbol{\gamma}_{\mathcal{B}} + \sum_{y \neq x} \boldsymbol{\gamma}_{\mathcal{A}_y} \right] \\
    g_{AB, pqrs} &= \braket{\phi_{AB, p}^{\mathrm{(HF)}}(\mathbf{r}_1) | \braket{\phi_{AB, q}^{\mathrm{(HF)}}(\mathbf{r}_2) | \Hat{g}_{AB}(1, 2) | \phi_{AB, s}^{\mathrm{(HF)}}(\mathbf{r}_2)} | \phi_{AB, r}^{\mathrm{(HF)}}(\mathbf{r}_1)} \\
    \Hat{g}_{AB}(1, 2) &= \frac{1}{|\Hat{r}_1 - \Hat{r}_2|}.
\end{align}
The coefficient matrix $\mathbf{C}$ transforms from the original HF basis to the Schmidt basis
\begin{align}
    \phi_{\mathcal{AB}, i} &= \sum_{j=1}^{d_{AB}} \phi_{AB, j}^{\mathrm{(HF)}} C_{ji} .
\end{align}
\end{subequations}
By considering that 
\begin{align}
\begin{split}
    \gamma_{AB, ij} =& 
    2 \sum_{k\in \mathrm{core}} \delta_{ik}\delta_{kj}
    + \sum_{m=1}^{d_A} \Bigg( 2 \lambda_{m} \delta_{im_A}\delta_{m_A j}
    + 2 (1-\lambda_{m}) \delta_{im_{\mathrm{bath}}}\delta_{m_{\mathrm{bath}}j}  \\
    & + 2 \sqrt{\lambda_{m}}\sqrt{1-\lambda_{m}} (\delta_{im_A}\delta_{m_{\mathrm{bath}}j} + \delta_{im_{\mathrm{bath}}}\delta_{m_{A}j}) \Bigg), 
\end{split}
\end{align}
the density matrices for the Schmidt basis (obtained from a restricted HF calculation) can be found.
Here, $\{m_A\}_{m=1}^{d_A}$ mark the impurity orbitals, while $\{m_{\mathrm{bath}}\}_{m=1}^{d_A}$ mark the corresponding bath orbitals.
Every impurity orbital $m_A$ is pairwise entangled with the bath orbital $m_{\mathrm{bath}}$ (see Eq.~(\ref{eq:decomp})).
Finally, the double counting energy $E_{\mathrm{dc}}$ corrects for the Coulomb and exchange energy that is double counted through the effective one-particle integrals.

The system-wide energy is then the sum of the different subsystem contributions
\begin{subequations}
\begin{align}
    E_{{\mathcal{A}}\mathcal{B}} &= E_{\mathrm{nn}, AB} +  E_\mathrm{core} + \sum_{x=1}^{N_\mathrm{imp}}E_{\mathcal{A}_x} - E_{\mathrm{dc}} \\
    E_{\mathcal{A}_x} &= \frac{\braket{\Psi_{\mathcal{A}_x} | \Hat{H}_{\mathcal{A}_x} | \Psi_{\mathcal{A}_x}}}{\braket{\Psi_{\mathcal{A}_x} |  \Psi_{\mathcal{A}_x}}}.
\end{align}
\end{subequations}

Note that LASSCF~\cite{gagl19,gagl20,gagl22_b,gagl25_b} is also a multi-CAS ansatz that uses the Schmidt decomposition to construct active spaces.
However, it differs from eCAS in several key aspects:
In order to construct the initial extended impurities, LASSCF assumes that the wave function can be represented as a product state of impurity states $\ket{\Psi_{A_x}}$.
This allows for the direct construction of impurity-specific bath orbitals.
Furthermore, once the extended impurities have been constructed, LASSCF performs a CASSCF calculation within each extended impurity, using the impurity orbitals as the initial active orbitals.
The impurity method is therefore necessarily a CASSCF method.
The CASSCF calculation admixes impurity orbitals with their corresponding bath orbitals, and upon convergence of the CASSCF calculations, the reduced environment is updated and the Schmidt decomposition is repeated.
Hence, LASSCF iteratively improves on the definition of the impurity and bath orbitals, rather than adding orbitals from the reduced environment to the active spaces.

\subsection{eCAS Embedding Protocol}
Below, the eCAS protocol is summarized:
\begin{enumerate}
    \item Calculate the system-wide wave function $\Psi_{AB}^\mathrm{(HF)}$ with HF.
    \item Select $d_A$ initial impurity orbitals $\{\phi_{A, i}\}_{i=1}^{d_{A}}$.
    \item Construct the extended impurity $\mathcal{A}$ and the reduced environment $\mathcal{B}$ via a Schmidt decomposition and localize the orbitals of the reduced environment.
    \item Calculate the mutual correlation matrix $\boldsymbol{\mathcal{M}}$ using a system-wide correlated wave function calculation. 
    \item Partition the extended impurity $\mathcal{A}$ into smaller extended impurities $\{\mathcal{A}_x\}_{x=1}^{N_\mathrm{imp}}$.
    \item Transfer all orbitals where $\max_{i\in\mathcal{A}_x}\mathcal{M}_{i, j\in\mathcal{B}} \geq \kappa$ from $\mathcal{B}$ to $\mathcal{A}_x$.
    If $\max_{i\in\mathcal{A}_x}\mathcal{M}_{i, j\in\mathcal{B}}$ and $\max_{i\in\mathcal{A}_{y\neq x}}\mathcal{M}_{i, j\in \mathcal{B}}$ are both larger than $\kappa$, $j$ is transferred to whichever impurity it has the largest mutual correlation with.
    \item Using $\boldsymbol{\mathcal{M}}$, calculate the quality estimate $\mathcal{Q}$.
    \item Set up and solve the multi-CAS problem.
\end{enumerate}

\section{Computational Methodology} \label{sec:computational_methodology}
We implemented eCAS in Python built on the PySCF environment~\cite{chan20_b}.
The constrained optimization problem of partitioning the impurity orbitals was solved with PySCIPOpt~\cite{serr16}.
The initial impurity orbitals were orthonormalized intrinsic atomic orbitals (IAOs)~\cite{kniz13}.
The reference minimal basis was MinAO~\cite{kniz13}.
Orthonormalization was achieved by a symmetric Löwdin orthonormalization.
In order to embed an atom, all IAOs of the atom were used as initial impurity orbitals $\{\phi_{A, i}\}$.
Localization of the occupied rotated environment orbitals was performed with the intrinsic bond orbital (IBO) scheme~\cite{kniz13}, while localization of the virtual rotated environment orbitals was performed through Cholesky factorization~\cite{koch06}.
The mutual correlation matrix $\boldsymbol{\mathcal{M}}$ was obtained from a Møller--Plesset perturbation theory calculation, truncated at second order (MP2).
The autoCAS program~\cite{reih19,reih24} was used to obtain threshold plots.

Cartesian coordinates of benzaldehyde were taken from PubChem~\cite{bolt19}.
Cartesian coordinates of hexanal, iminohexanal (6-iminohexanal), iminooctanal (8-iminooctanal) and iminodecanal (10-iminodecanal) were obtained after structure optimization with restricted HF in the 6-31G basis set.
For all of the investigated molecular systems, we deliberately use the small 6-31G atomic basis set~\cite{popl71,popl72}, because it 
allows for easy visualization and computationally fast system-wide reference calculations with the impurity method.
However, we emphasize that eCAS will also perform well for decent atomic basis sets that are well suited for ab initio correlation methods.
We solve the active space problem using CCSD, meaning that the impurity method is CCSD, while the environment method is restricted HF, which we denote as eCAS(CCSD/HF).
Singlet excitation energies are calculated with equation-of-motion (EOM) CCSD, as implemented in PySCF.

We compare eCAS(CCSD/HF) to active space calculations with natural orbitals.
Thereby, the $d_{\mathrm{active}}$ natural orbitals with the occupation numbers that deviate the most from zero or two are considered as active.
This is a common strategy to construct active spaces and we follow the PySCF implementation of this procedure.
The natural orbitals are obtained from an MP2 calculation and the active space is solved using CCSD.
We abbreviate this active space method as MP2 AS.
Comparison to one-shot DMET and EWF was possible through calculations performed with Vayesta~\cite{boot22,boot23}.
EWF and DMET require the full molecule to be tiled into impurities.
This was accomplished by assigning each atom that is not part of the impurity of interest to its own separate impurity.
Because DMET must cover the full virtual space as well, IAOs and projected atomic orbitals (PAOs)~\cite{pula83} were used as the initial impurity orbitals $\{\phi_{A_x, i}\}_{i=1}^{d_{A_x}}$.
In one-shot DMET, the impurity one-body potentials were set to zero and not iteratively fitted.
However, the global chemical potential was still self-consistently optimized.
For EWF, the chemical potential was not converged and a one-shot procedure was applied instead.
This led to considerable computational savings and it has been shown that the results are not adversely affected~\cite{boot22}.
We used the definition of Vayesta to select the bath natural orbital occupation number criterion $\eta$ for EWF.
Similarly to $\kappa$, $\eta$ governs how many orbitals will be transferred from $\mathcal{B}$ to $\mathcal{A}$.
For both EWF and DMET we chose CCSD as the impurity method.

\section{Results and Discussion}

\subsection{Walk-Through of eCAS}
In this section, we present the eCAS methodology applied to benzaldehyde.
In the 6-31G basis set and with the aldehydic hydrogen atom as the impurity, it holds that $d_{AB} = 84$, $d_A = 1$, $d_B = 83$ and $n_{AB} = 28$.

\begin{figure}
    \centering
    \includegraphics[width=\linewidth]{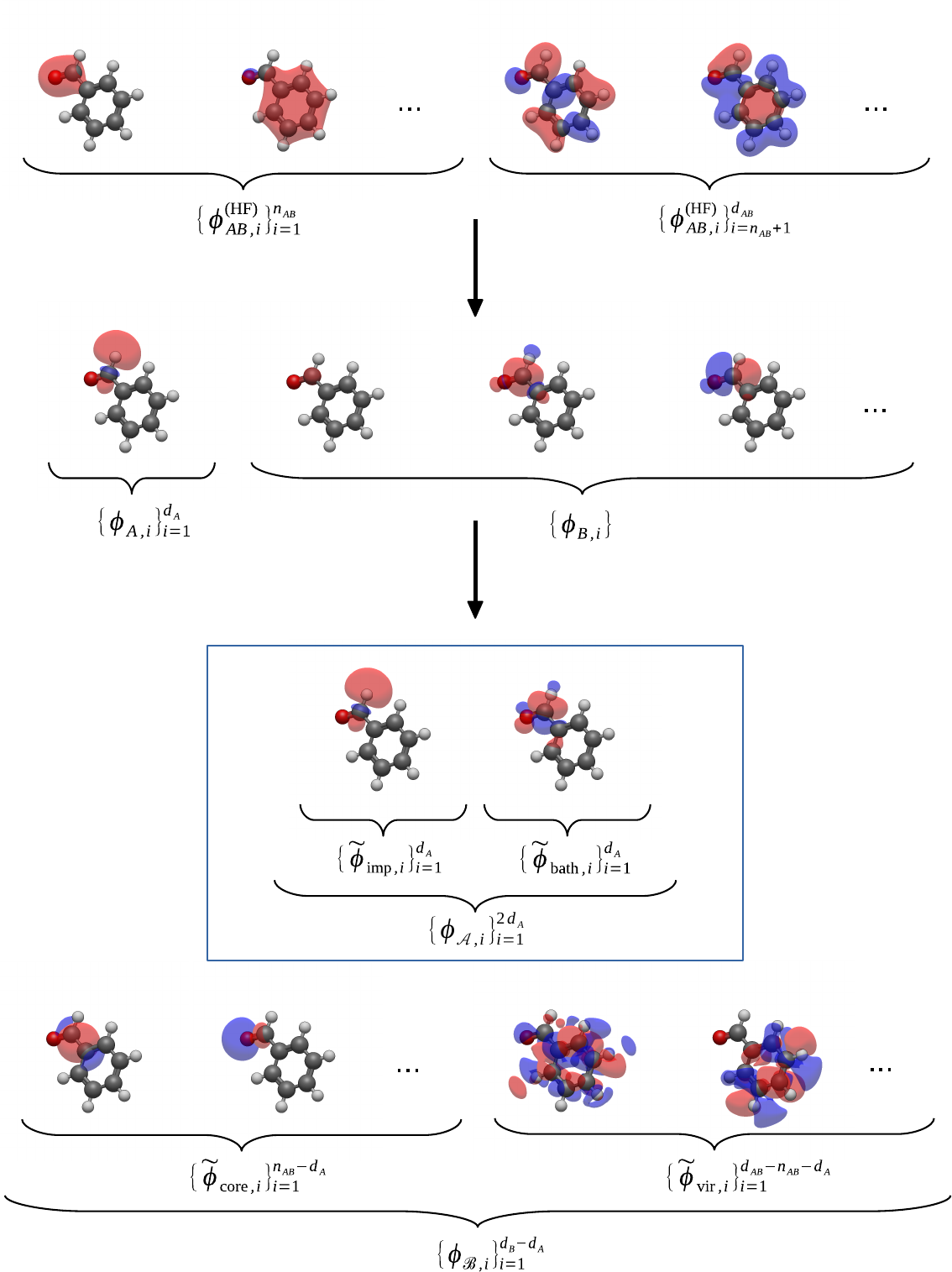}
    \caption{Construction of the Schmidt basis $\{\phi_{\mathcal{AB}, i}\}_{i=1}^{d_{AB}}$ for benzaldehyde. Initially, the canonical HF orbitals are calculated. Next, IAOs are constructed and $d_A$ impurity IAOs are selected, which are subsequently used to construct the Schmidt basis. Orbitals of the extended impurity $\mathcal{A}$ are highlighted with a blue border.}
    \label{fig:1}
\end{figure}

In figure~\ref{fig:1}, we show the construction of the Schmidt basis.
First, a system-wide restricted HF calculation is performed, yielding canonical HF orbitals $\{\phi_{AB, i}^\mathrm{(HF)}\}_{i=1}^{d_{AB}}$, $n_{AB}$ of which are doubly occupied.
In the next step, the HF orbitals are used to construct orthonormalized IAOs.
$d_A$ impurity IAOs $\{\phi_{A, i}\}_{i=1}^{d_A}$ are then selected and used to rotate and project the HF orbitals to the rotated orbitals $\{\Tilde{\phi}_{AB, i}\}_{i=1}^{d_{AB}}$.
eCAS does not preserve the initial choice of impurity orbitals $\{\phi_{A, i}\}_{i=1}^{d_A}$.
However, because of the projection into the impurity space, they are part of the same subspace as the initial impurity IAOs.
The bath orbitals are constructed such that they capture the mean-field entanglement between impurity and environment exactly.
As a consequence, they remain closely localized near the impurity.

In eCAS, the impurity and bath orbitals are always part of the extended impurity $\mathcal{A}$, and thus of the active space.
Orbitals of the reduced environment $\mathcal{B}$ are, in contrast, only transferred to the extended impurity $\mathcal{A}$ if their mutual correlation with any of the impurity or bath orbitals is larger than the threshold $\kappa$.
To find the mutual correlation between all pairs the 1- and 2-RDMs of a correlated wave function (here an MP2 wave function) are calculated in the original HF basis and subsequently rotated into the Schmidt basis.
This has the advantage that no 4-index transformation is necessary.
Additionally, computational savings can be achieved by freezing energetically low-lying MOs and by density fitting.
For more correlated systems, an MP2 calculation will no longer adequately capture the mutual correlation and alternative methods, such as CCSD(T), will be necessary.
For systems that are strongly correlated and where a single-reference method is no longer adequate, a multi-configurational solver such as the density matrix renormalization group will be required.

\begin{figure}[H]
    \centering
    \includegraphics[width=\linewidth]{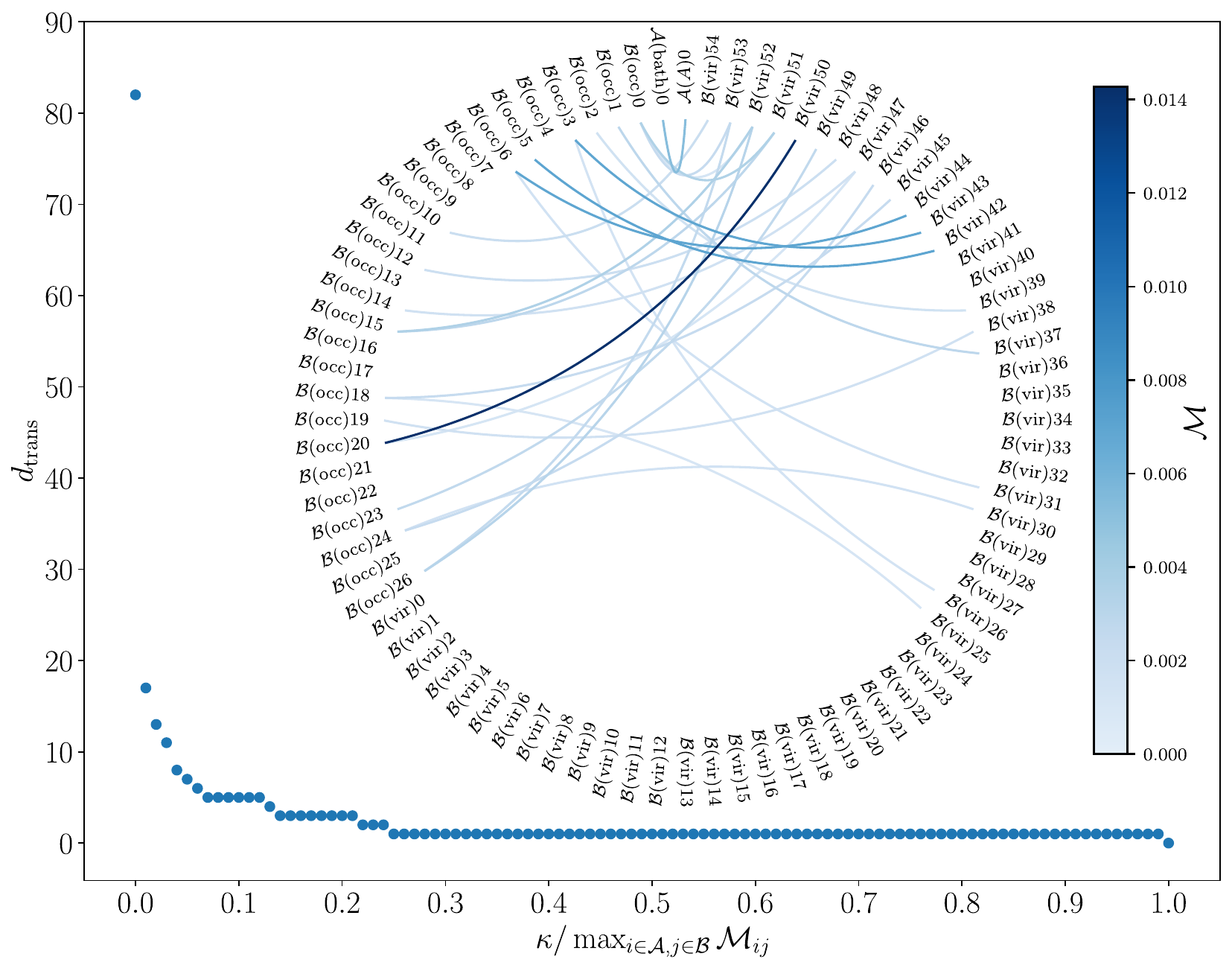}
    \caption{Threshold plot and orbital mutual correlation diagram (inset) of benzaldehyde. `$\mathcal{A}(A)$' marks the impurity orbital, `$\mathcal{A}(\mathrm{bath})$' the bath orbital. `$\mathcal{B}(\mathrm{occ})$' the 27 occupied reduced environment orbitals and `$\mathcal{B}(\mathrm{vir})$' the 55 virtual reduced environment orbitals. The number of orbitals $d_{\mathrm{trans}}$ that are transferred from the reduced environment $\mathcal{B}$ to the extended impurity $\mathcal{A}$ is governed by the threshold $\kappa$. In the threshold plot, $\kappa$ is normalized to the largest mutual correlation between $\mathcal{A}$ and $\mathcal{B}$.}
    \label{fig:2}
\end{figure}

Once the 1- and 2-RDMs have been obtained, the mutual correlation matrix $\boldsymbol{\mathcal{M}}$ can be calculated.
Thereby, further computational savings can be gained by the observation that $\mathcal{M}_{ij} = \mathcal{M}_{ji}$.
An orbital mutual correlation diagram~\cite{reih11,reih16,reih19} can provide insights into the correlation pattern of the electronic structure of a molecule.
In figure~\ref{fig:2} we see that impurity and bath orbital are correlated with a mutual correlation of $\mathcal{M} = 5.3 \cdot 10^{-3}$.
Considering the close spatial proximity of these orbitals, their orbital shape (see figure~\ref{fig:1}), and their entanglement, this can be expected.

The mutual correlation between extended impurity orbitals and reduced environment orbitals is, by contrast, considerably smaller and the orbital pair with the highest mutual correlation has $\max_{i\in \mathcal{A}, j\in\mathcal{B}} \mathcal{M}_{ij} = 4.5 \cdot 10^{-4}$.
Note that the construction of the full mutual correlation matrix is not necessary for the augmentation of the extended impurity, as we are only interested in $\mathcal{M}_{i\in\mathcal{A}, j\in\mathcal{B}}$.
This also means that the calculation of the full 1- and 2-RDMs is not necessary.

If we are only interested in how many orbitals will be transferred from $\mathcal{B}$ to $\mathcal{A}$ for a given threshold $\kappa$, we can consider the threshold plot~\cite{reih16,reih19} in figure~\ref{fig:2}.
There are several plateaus, where decreasing $\kappa$ will not change the number of transferred orbitals.
In autoCAS~\cite{reih16,reih19}, these plateaus are used to choose the active space.

\begin{figure}[H]
    \centering
    \includegraphics[width=\linewidth]{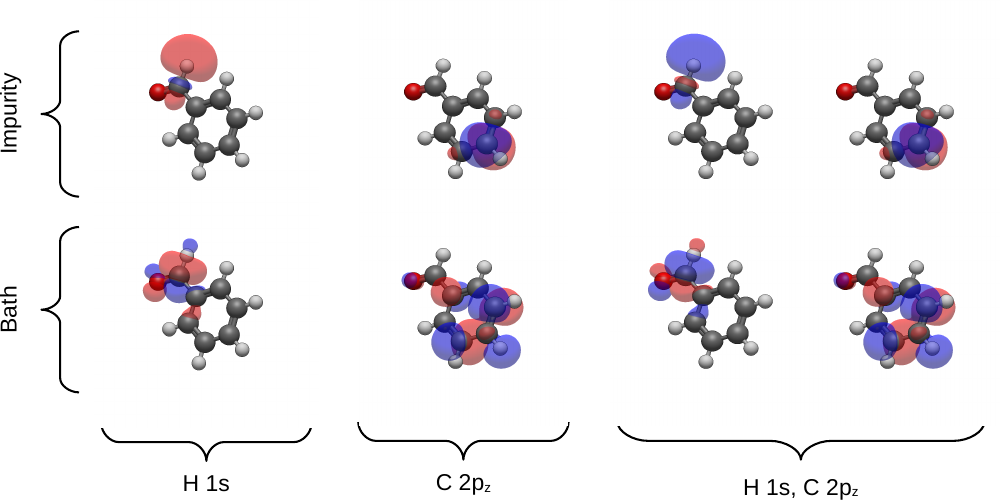}
    \caption{Impurity and bath orbitals for different selections of initial impurity orbitals $\{\phi_{A, i}\}$. The leftmost column uses the aldehydic hydrogen $1s$ orbital as the initial IAO impurity orbital, while the second column uses the $2p_z$ orbital of the carbon in \textit{para}-position to the aldehyde group. The rightmost two columns show the impurity and bath orbitals if both the hydrogen $1s$ and the carbon $2p_z$ orbitals are chosen as initial impurity orbitals.}
    \label{fig:3}
\end{figure}

eCAS only constructs one Schmidt basis, even in the case of multiple impurities.
For more than one impurity this could lead to scenarios where impurity and bath orbitals cannot be clearly assigned.
However, as long as the different impurities are far enough apart, these scenarios typically do not occur.
To demonstrate this, we show the impurity and bath orbitals for benzaldehyde for the case that we choose a second impurity: the $2p_z$ orbital of the carbon in \textit{para}-position to the aldehyde group.
From figure~\ref{fig:3} it becomes apparent that the extended impurity orbitals of the joint impurities are unchanged compared to when the Schmidt basis is built for the individual impurities.

\subsection{Comparison to Alternative Methods}
\begin{figure}
    \centering
    \includegraphics[width=\linewidth]{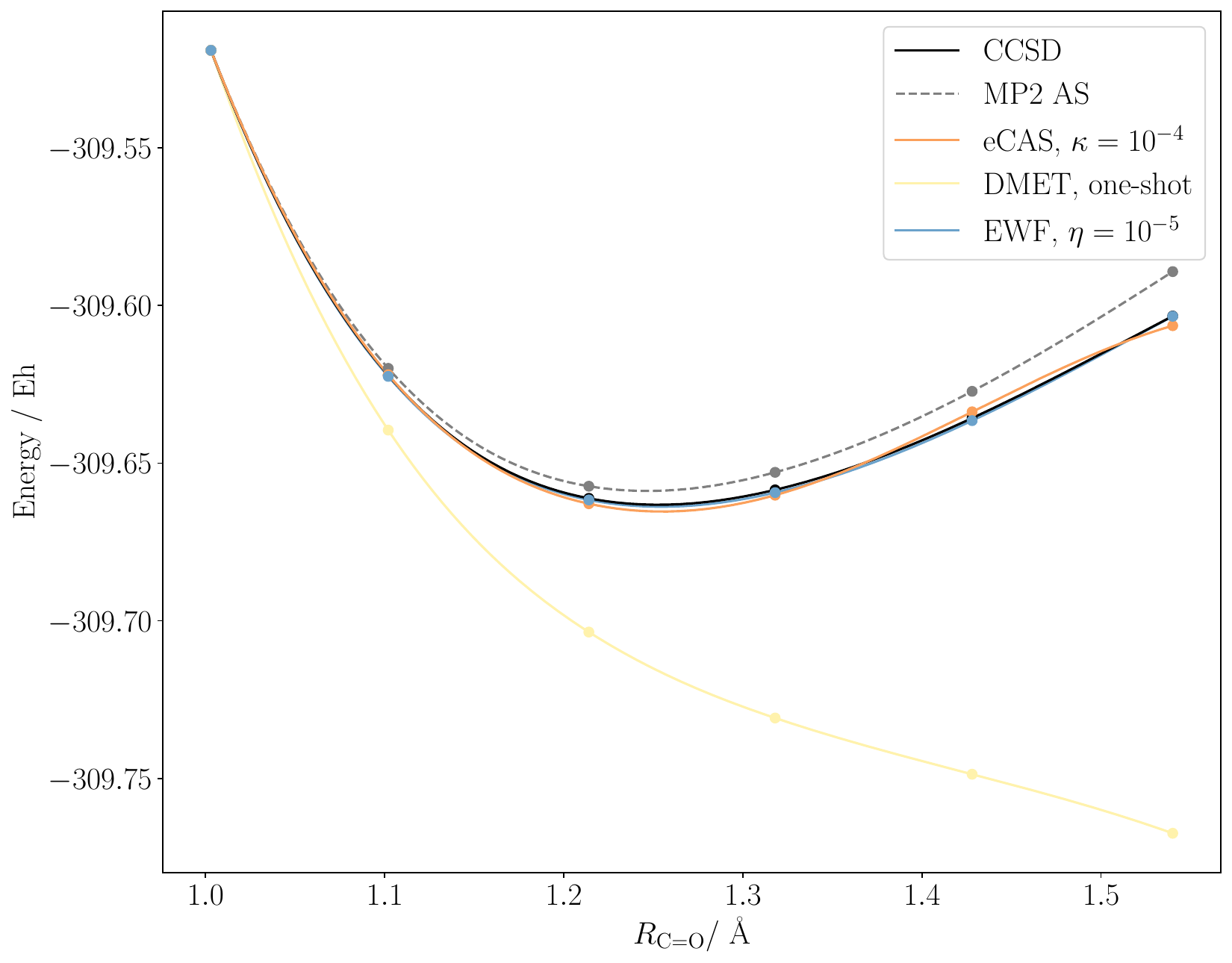}
    \caption{Electronic energy of hexanal as a function of the carbonyl bond length. eCAS, EWF, DMET, and MP2 AS all include approximately 40\% of the orbitals in the (largest) impurity. Note that EWF and DMET require the system to be fully fragmented into atomic impurities, except for C=O, which is one fragment.}
    \label{fig:4}
\end{figure}

We now investigate a potential energy surface (PES) scan of hexanal along the carbonyl stretching coordinate.
To this end, we use all IAOs of the carbonyl carbon and oxygen atoms as initial impurity orbitals.
Figure~\ref{fig:4} shows that eCAS(CCSD/HF) gives comparable results to when the full molecule is treated with CCSD.
At $\kappa=10^{-4}$, eCAS includes 36 out of 87 orbitals for all points calculated on the PES.

Constructing the active space from $d_{\mathrm{active}} = 36$ MP2 natural orbitals leads to worse results.
In figure~\ref{fig:4} we also compare the results to two other embedding methods, namely one-shot DMET and EWF.
DMET, with its use of IAOs/PAOs as initial impurity orbitals, consistently has 27 orbitals in the largest active space.
EWF with $\eta=10^{-5}$ consistently has 38 orbitals in the largest active space.
EWF almost perfectly coincides with the CCSD results, while one-shot DMET is unable to accomplish this.
While we assume that iteratively converged DMET performs better, we could not test this as these calculations did not converge.

Comparing the computational timings, eCAS, EWF, and MP2 AS are faster than CCSD, while one-shot DMET is slower.
This is because one-shot DMET also refines the global chemical potential.
Table~\ref{tab:cmp} compares the different features of eCAS, DMET and EWF.

\begin{table}
    \caption{Conceptual comparison of eCAS, DMET, and EWF. `$E_A$ only' indicates that an energy calculation of the impurity with the impurity method suffices. `Iterative' indicates whether an iterative evaluation of the impurity with the impurity method is needed, and `$\Psi_{AB}^\mathrm{(corr.)}$' whether a system-wide correlated wave function must be evaluated. `Impr.' indicates whether a method is systematically improvable and `Frag.' whether the method can work as a fragmentation method, where the full molecule is fragmented and each fragment is considered as an impurity. `y' indicates yes, `n' indicates no.}
    \label{tab:cmp}
    \centering
    \begin{tabular}{c|c|c|c|c|c}
    \hline\hline
    Method  & $E_A$ only    & Iterative & $\Psi_{AB}^\mathrm{(corr.)}$  & Impr.  & Frag. \\ \hline
    eCAS    & y             & n         & y                             & y     & n     \\
    DMET    & n             & y         & n                             & n     & y     \\
    EWF     & n             & n         & y                             & y     & y     \\\hline\hline
    \end{tabular}
\end{table}

In its current formulation, eCAS slightly deviates in the PES scan of hexanal from EWF.
We assume that this is mainly due to an inferior localization of the reduced environment orbitals in eCAS.
While EWF localizes the environment orbitals in a scheme similar to the construction of natural orbitals, eCAS uses IBO and Cholesky localization, which typically leads to more delocalized orbitals.
With an improved localization procedure, a smaller number of reduced environment orbitals would capture a larger amount of mutual correlation to the extended impurity.
Hence, $d_{\mathrm{trans}}$ would be smaller for the same $\kappa$.

\subsection{Assessment of the Quality Measure}
We now investigate the efficacy of the quality measure $\mathcal{Q}$.
To this end, we study the two lowest-lying excitations in iminohexanal, iminooctanal and iminodecanal.
The two excitation energies are dominated by the $n \to \pi^*$ transitions of the imine and the aldehyde groups.
We consider the $2p_z$ orbitals of the carbonyl carbon and the carbonyl oxygen as the initial impurity orbitals of the first impurity ($\{\phi_{A_1, i}\}_{i=1}^2$), and the $2p_z$ orbitals of the imine carbon and the imine nitrogen as the initial impurity orbitals of the second impurity ($\{\phi_{A_2, i}\}_{i=1}^2$).
The character of the non-bonding orbitals is included in the impurities through bath and transferred environment orbitals.
The excitation energies are then calculated separately for the two extended impurities using singlet EOM-CCSD.

\begin{figure}[H]
    \centering
    \includegraphics[width=\linewidth]{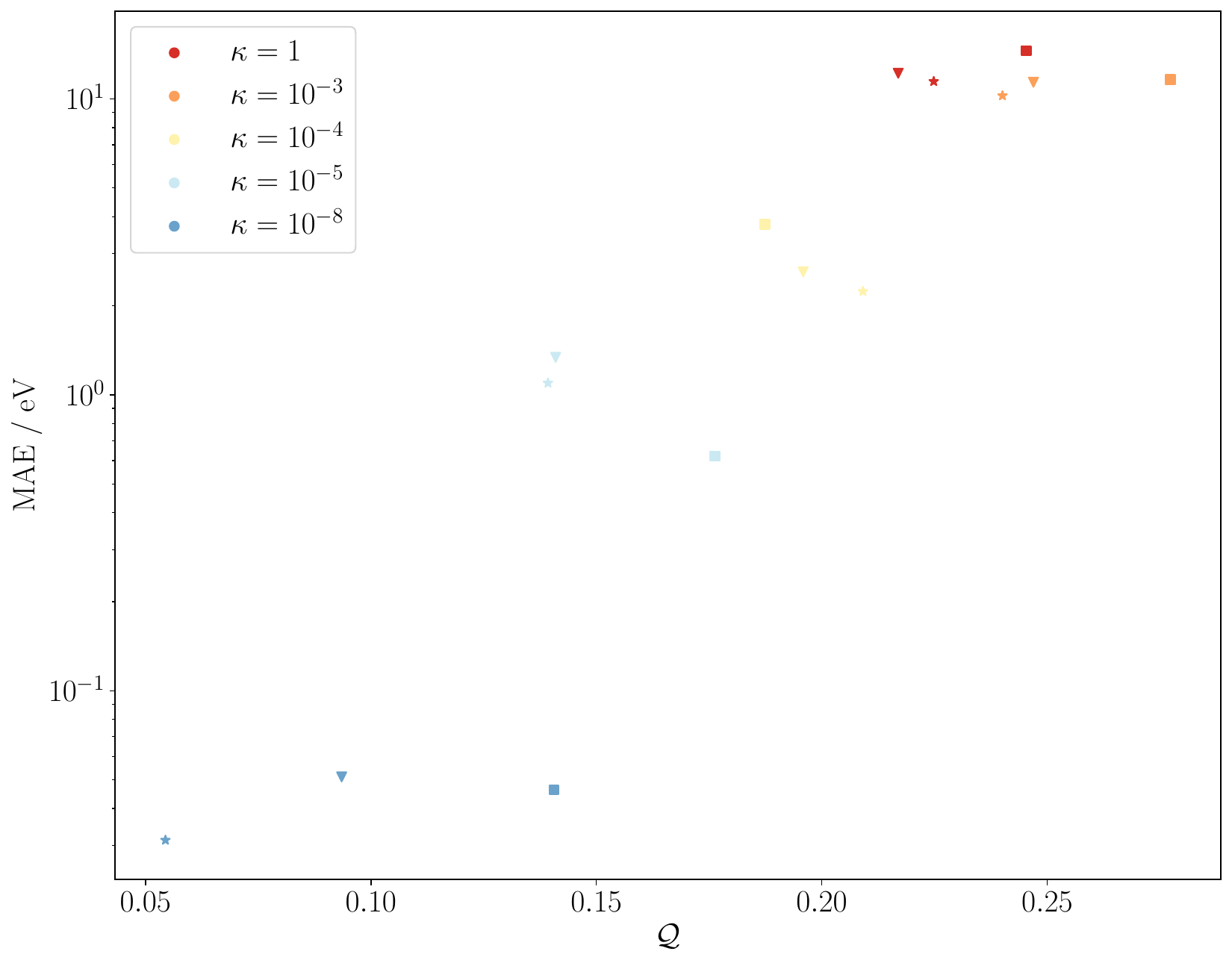}
    \caption{Mean absolute error of the first and second excitation energy of iminoalkanals. Note the relationship between the error and the quality estimate $\mathcal{Q}$. Stars mark iminohexanal, triangles iminooctanal, and squares iminodecanal.}
    \label{fig:5}
\end{figure}

The mean absolute error (MAE) of the first and second excitation energies of iminohexanal, iminooctanal and iminodecanal and the associated quality measure $\mathcal{Q}$ are shown in figure~\ref{fig:5}.
The MAE is relative to the first two excitation energies of a full-system singlet EOM-CCSD calculation.
Decreasing $\kappa$ improves the MAE, as it also increases the size of the extended impurity.
This does not, however, mean that $\mathcal{Q}$ is systematically lowered (consider that lowering $\kappa$ from $1$ to $10^{-3}$ increases $\mathcal{Q}$).
This is because there are now more inter-subsystem pairs that contribute to $\mathcal{Q}$.
Note that for iminohexanal $\kappa=1$ ($\kappa = 10^{-8}$) includes approximately 4.3\% (50\%) of the orbitals in each impurity, while for iminodecanal it includes only approximately 2.7\% (29.5\%).
$\kappa = 1$ means that no orbitals were transferred ($d_\mathrm{trans} = 0$).

\section{Conclusions}
\label{sec:conclusion}
In this work we introduced the eCAS quantum embedding protocol. We designed eCAS in such a way that it is well-suited for quantum computing applications:
It neither requires the calculation of 1- or 2-RDMs with the impurity method, nor does it need iterative evaluations of the impurity description.
In addition, eCAS is systematically improvable and possesses a quality measure to control its reliability.

Despite these merits of eCAS for quantum computing,
eCAS will also prove useful for embedding problems on classical computers.
It can be viewed as a natural extension of the autoCAS algorithm~\cite{reih16,reih19}, 
where we now extend the question `what is the best active space?' by the next challenge `given an active space, what are the best orbitals to add to it?'.
Finding the solution to the latter question could be instrumental in improving active spaces that were selected by chemical intuition or by using orbital preparation schemes such as AVAS.

By studying the first and second excitation energy of iminoalkanals, we could show that lowering the threshold $\kappa$ systematically improves the accuracy of eCAS, irrespective of the size of the iminoalkanal.
Thereby, the quality measure $\mathcal{Q}$ can be used as an indicator of the error incurred by eCAS.
We have also compared the capabilities of eCAS to one-shot DMET, EWF, and results of a traditional active space calculation.
This comparison was based on electronic energies of hexanal as a function of the carbonyl bond length.
While one-shot DMET and the active space built from MP2 natural orbitals were unable to adequately recover the electronic energies, both eCAS and EWF proved to be reliable.

Since eCAS simply prepares the orbitals for a multi-CAS calculation, any CAS-based solver can be chosen as impurity method.
This facilitates the calculation of analytical gradients and molecular properties, which we will consider in forthcoming work, and it allows for the simple application of algorithms such as direct orbital selection (DOS) procedures~\cite{neug19_b,neug20,reih23,neug24_b}, which is key for defining consistent active spaces along chemical reaction coordinates.
Similarly, eCAS can be easily extended toward CASSCF-type models plus perturbative corrections.

\section*{Acknowledgments}
\label{sec:acknowledgments}
This work was financially supported by the Swiss National Science Foundation (Grant No. SNF 200021\_219616).

\providecommand{\latin}[1]{#1}
\makeatletter
\providecommand{\doi}
  {\begingroup\let\do\@makeother\dospecials
  \catcode`\{=1 \catcode`\}=2 \doi@aux}
\providecommand{\doi@aux}[1]{\endgroup\texttt{#1}}
\makeatother
\providecommand*\mcitethebibliography{\thebibliography}
\csname @ifundefined\endcsname{endmcitethebibliography}
  {\let\endmcitethebibliography\endthebibliography}{}

\end{document}